\documentclass[aps,prl,reprint,superscriptaddress]{revtex4-1}
\usepackage{graphicx}
\usepackage{color}
\usepackage{epsfig}
\usepackage[colorlinks]{hyperref}

\begin{document}
\title{ Dipole Alignment of Water Molecules Flowing Through Carbon Nanotube } 
\author{Hemant Kumar}
\email{hemant@iitbbs.ac.in}
\author{Saheb Bera}
\affiliation{School of Basic Sciences, Indian Institute of Technology Bhubaneswar, Argul, Odisha 752050, India}
\author{Subhadeep Dasgupta}

\author{A. K. Sood}
\affiliation{Department of Physics, Indian Institute of Science, Bangalore, 560012, India}
\author{Chandan Dasgupta}
\affiliation{Department of Physics, Indian Institute of Science, Bangalore, 560012, India}
\affiliation{International Centre for Theoretical Sciences, TIFR, Bangalore 560089, India}

\author{Prabal K. Maiti}
\email{maiti@iisc.ac.in}
\affiliation{Department of Physics, Indian Institute of Science, Bangalore, 560012, India}


\begin{abstract}
The fast flow rate of water through nanochannels has promising applications in desalination, energy conversion, and nanomedicine. We have used molecular dynamics simulations to show that the water molecules  passing through a wide single-walled carbon nanotube (CNT) cavity get aligned by flow to have a net dipole moment along the flow direction. With increasing flow velocity, the net dipole moment first increases and eventually saturates to a constant value. This behavior is similar to the Langevin theory of paraelectricity with the flow velocity acting as an effective aligning field. We show conclusively that the microscopic origin of this behavior is the preferential entry of water molecules with their dipole vectors pointing inward along the CNT axis.
\end{abstract}

\maketitle
Recent years have seen an upsurge of interest in exploring ultrafast transport of water in various nanochannels with potential applications such as desalination, separation process, and energy conversion~\cite{bocquet2020nanofluidics,xie2018fast, shen2021artificial, doi:10.1021/acs.nanolett.9b04552, doi:10.1021/acsomega.0c05584,doi:10.1021/nl1021046,siria2013giant,heiranian2020revisiting,chen2013nature,davis2020pressure,zhang2013water}. However, a quantitative understanding of flow-induced effects in nanochannels is still lacking~\cite{doi:10.1021/acsnano.9b04328,PhysRevLett.101.257801}. Recent progress in nanofabrication and the development of measurement techniques have enabled direct investigation of fluids confined to nanochannels~\cite{li2017direct, doi:10.1021/acs.accounts.9b00411,agrawal2017observation, doi:10.1021/acsnano.0c08634, verhagen2020anomalous}. Tunuguntla et {\it al.} demonstrated a high permeability of water across narrow carbon nanotubes of diameter $\sim 0.8 $nm and significantly lower for wider CNTs (diameter $\sim 1.5$ nm)\cite{tunuguntla2017enhanced}. Radha et {\it al.} observed relatively fast flow rates when a slit pore is narrow to accommodate only a monolayer of water molecules \cite{radha2016molecular}. The flow across the nanochannels can also be modulated by applying static or oscillating electric fields, indicating a strong correlation between the dipole ordering and the flow rate~\cite{mouterde2019molecular,doi:10.1021/nl203614t,doi:10.1021/acs.jpclett.9b03228,wang2020net,zhang2020electrokinetic}. However, such an association of structural ordering and flow rate is not demonstrated yet~\cite{wei2014understanding}. Yuan et {\it al.} have shown, using ab-initio molecular simulation, that the net dipole moment of water molecules inside a carbon nanotube can polarize the tube, thus creating a potential difference across the nanotube ends\cite{zhao_jacs_2009}. Ghosh et {\it al.} have shown experimentally the generation of voltage (and current) across the CNT due to the pressure-induced flow of water molecules across nanotube bundles\cite{sood_sci}. This observation motivated us to ask if water dipoles order when water flows inside the CNTs. The answer indeed is in the affirmative as shown in our present molecular dynamics (MD) simulations. 

Water molecules inside narrow carbon nanotubes (diameter $\sim 0.8$ nm) in equilibrium are arranged in a single file manner, with all the dipoles pointing in the same directions along the nanotube axis \cite{mukherjee:124704, chakraborty2017confined}.
 These dipoles flip collectively on a time scale that depends on the length of the nanotube\cite{doi:10.1021/nn800182v}. Lin et {\it al.} have shown that in equilibrium, the dielectric relaxation of confined water molecules inside CNTs is much faster along the cross-section but slower along the axis\cite{PhysRevB.80.045419}. Inside narrow CNTs, the dipole vector of the water molecules prefers to stay along the axis while it relaxes faster perpendicular to the nanotube axis. However, as the diameter of the nanotube increases, water molecules are no longer arranged in a single file manner, and there is no apparent ordering at ambient conditions of pressure and temperature.

In this paper, we demonstrate that the water dipoles confined inside a wider (10, 10) nanotube (diameter 1.4 nm), which exhibit no net dipole polarization in equilibrium, can show net dipole ordering under the effect of flow generated by a pressure gradient. The net alignment of dipoles along the flow direction increases with flow velocity, saturating to a finite value. This behavior of confined water can be very well explained by treating them as independent dipoles under some effective aligning field that depends on the flow velocity. The electric field stems from the asymmetric dielectric interface at the CNT ends where water molecules from the bulk enter the CNT cavity. This observation not only demonstrate the existence of a novel physical phenomenon but also suggests pathways for various nano-electromechanical devices.

\begin{figure}
\includegraphics[width=\linewidth]{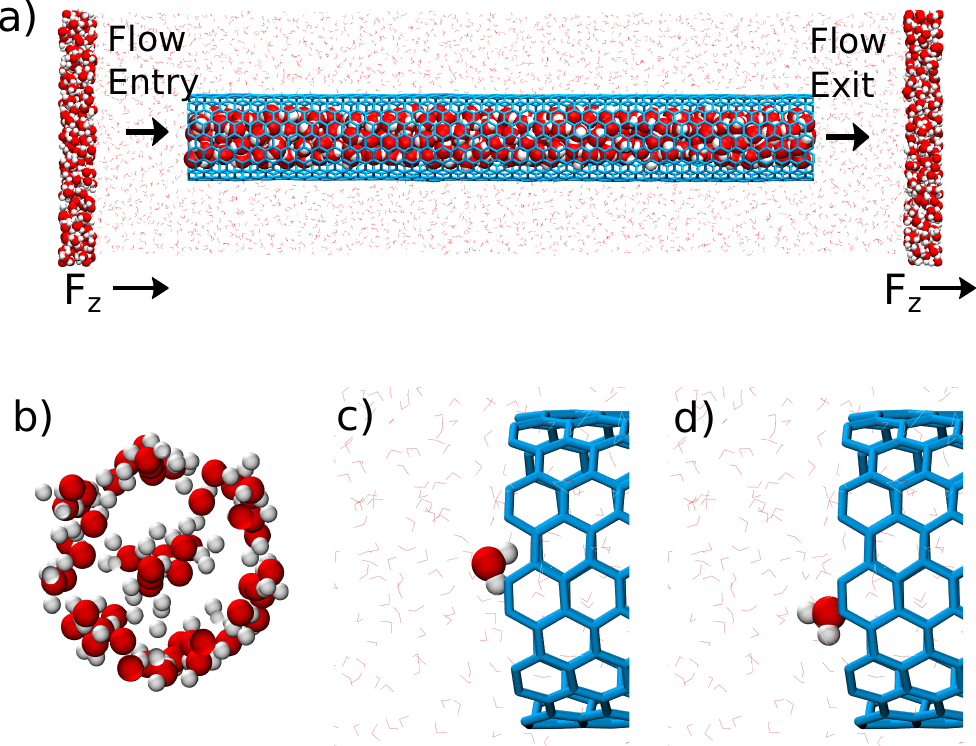}
\caption{ Figure 1: {\bf a) }  Simulated system, highlighted water molecules at both ends represent the layers on which additional force was applied to simulate the flow. {\bf b) }  Structure of water molecules confined inside the carbon nanotube. A typical water molecule before entering CNT with dipole pointing inward (panel {\bf c}) and outward (panel {\bf d}).}
\label{simbox}
\end{figure}

We have performed MD simulations of a 10 nm long armchair (10, 10) carbon nanotube (diameter 1.4 nm) solvated in TIP3P~\cite{jorgensen1983comparison} water as shown in Fig. \ref{simbox}a   using LAMMPS~\cite{lammps}. Carbon atoms were modeled as uncharged Lennard-Jones (LJ) particles with parameters taken from AMBER force-field~\cite{PMID:15116359,majumdar2021dielectric}. Interaction between water and carbon atoms was included through LJ interaction between carbon and oxygen atoms. Long-range electrostatic interactions between charged entities were computed using PPPM algorithm with the real space cut-off of 10 \AA~\cite{lammps}.  We initially run equilibrium MD simulation in NPT keeping temperature at 280 K and pressure P = 1 atm. Nose-Hoover thermostat and barostat as implemented in LAMMPS were used to maintain the constant temperature and pressure\cite{PhysRevB.69.134103,doi:10.1063/1.467468}. After equilibrating the system at T = 280 K and P = 1 atm, we switched to NVT ensemble using Nose-Hover  thermostat with a coupling constant of 1 ps. To simulate pressure gradient induced flow, we pushed a 5 \AA  layer of water on both the ends of the water-box along the nanotube axis by adding extra force on the oxygen atoms of each water molecule lying in these layers [Fig \ref{simbox}a]. This protocol for simulating hydrostatic pressure gradients is known to mimic pressure-induced flow~\cite{ISI:000176445800014}. We waited for 5 ns of simulation time to achieve a steady state. Later, 50 ns of flow simulation was used to analyze the behavior of water dipoles at different flow velocities.

\begin{figure*}
\includegraphics[width=\linewidth]{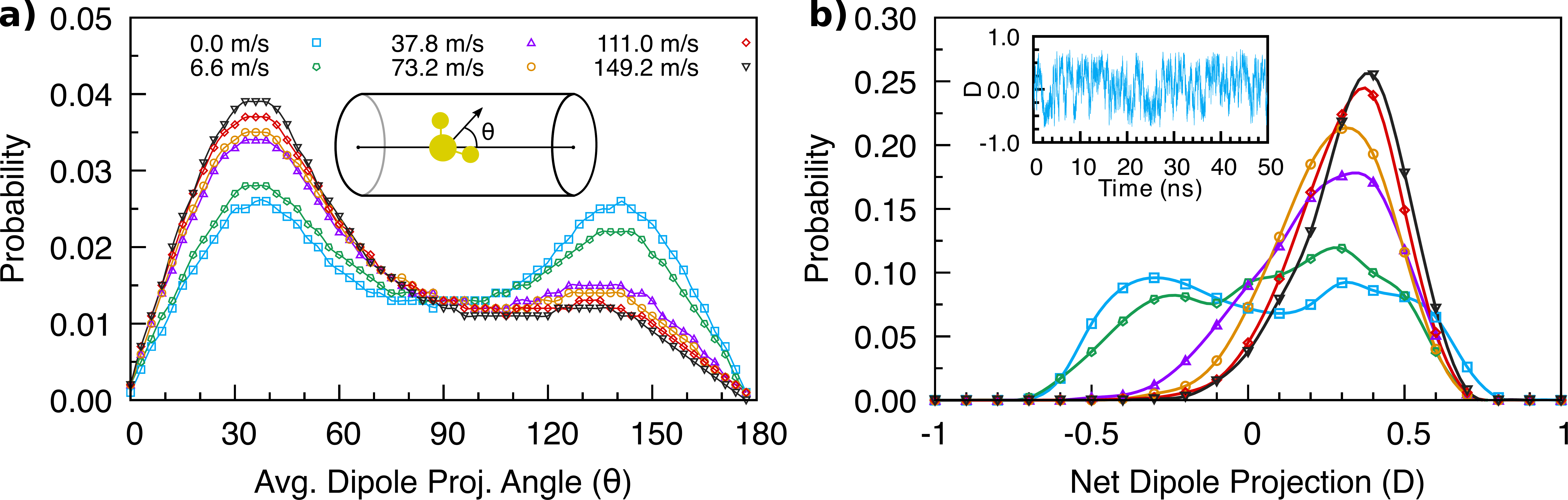}  
\caption{{\bf a) }Distribution of the angle between the dipole moment of individual water molecules inside a (10, 10) carbon nanotube and the flow direction. In the absence of flow, the probability shows peaks of equal height at  $\theta=40^{\circ}$ and $\theta=140^{\circ}$ . Flow creates asymmetry, and more dipoles are orientated along the flow direction, {\bf b)} Probability distribution of the net dipole projection of all confined water molecules for various flow velocities. Inset shows the temporal variation of the net dipole projection of confined water molecules along the CNT axis in equilibrium.}
\label{dipole_dist}
\end{figure*}
At equilibrium, when there is no flow, water dipoles spend an almost equal amount of time aligned either parallel or antiparallel to the nanotube axis, and the average dipole moment of the confined water molecules over a long trajectory is zero. Fig. 2a shows the probability of finding a dipole in a particular alignment for different flow velocities. Equal height peaks at $\theta=40^{\circ}$ and $\theta=140^{\circ}$   in equilibrium demonstrate the equal preference for both the alignments. At finite flow velocity, more dipoles prefer to be aligned along the flow direction. Fig. 2a clearly brings out that the probability of alignment along the flow direction (+z axis) at $\theta=40^{\circ}$ increases while alignment opposite to the flow direction (peak at  $\theta=140^{\circ}$  ) (-z axis) decreases, leading to a net dipole moment along the direction of flow. 
\begin{figure}[htbp]
\includegraphics[width=\linewidth]{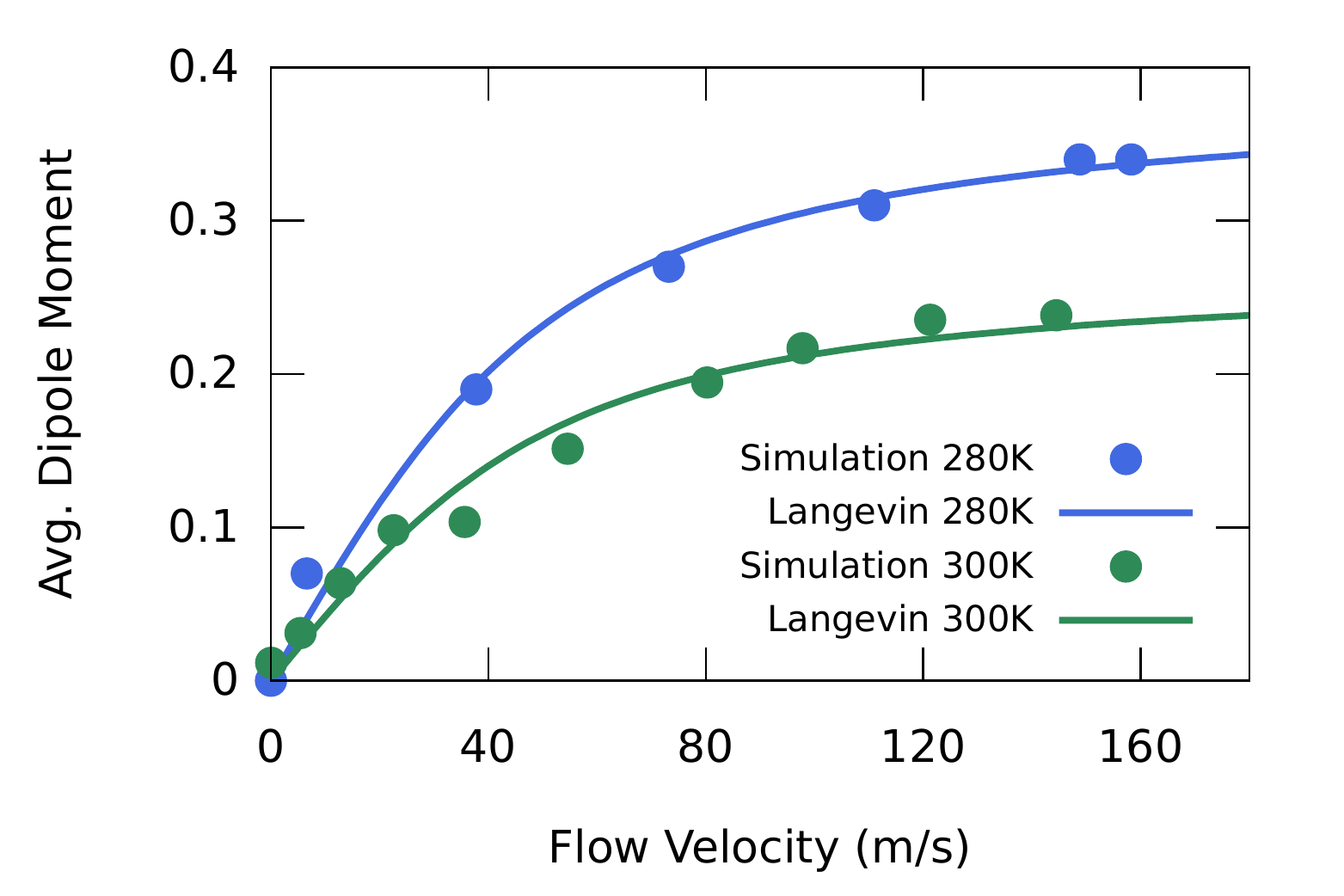}
\caption{ Variation of the average dipole moment per molecule of the confined water molecules, projected in the flow direction, with the flow velocity for temperatures T = 280 K and 300 K. The variation can be fitted well with the Langevin function (Eq. 1, $D_0=0.39$ and $a=13.28 K/m.s^{-1}$ for T = 280 K and $D_0=0.27$ and $a=13.28 K/m.s^{-1}$ for T = 300 K), which implies that the flow effect can be well described by an effective field created due to the flowing water molecules.  As shown in Fig. S3, the Langevin description works for CNTs of different length and the average dipole moment increases with the CNT length. Variation of the net dipole moment of the confined water molecules with flow velocity at three different lengths. }
\label{lang_fit}
\end{figure}
To investigate the nature of this alignment, we study the spatial ordering of confined water molecules for different flow velocities by calculating the pair-correlation function $g(z)$ of the confined water molecule using  $g(z)=\displaystyle\sum\limits_{i=1}^N\sum\limits_{j> i}^N \langle \frac{\delta(z-z_{ij})}{N}\rangle$, where $N$ is the number of water molecules inside the nanotube in each snapshot of the system and  $z_{ij}$ is the separation along the nanotube axis between $i^{th}$  and  $j^{th}$ water molecules. The computed pair correlation functions for the confined water molecules at different flow velocities demonstrate that $g(z)$ is almost independent of flow velocities with a very slight increase in the ordering at higher flow velocities (Fig. S1). This implies that flow does not alter positional ordering, but its net effect is to create an aligning field that tends to orient the dipoles of confined water molecules in the direction of flow.
At a given temperature, increasing the flow velocity increases the net dipole moment along the flow direction, and then it saturates to a constant value. Fig. \ref{lang_fit} shows the variation of the average projection of the dipole moment vector of the confined water molecules for two different temperatures. The value of the saturated dipole moment depends on the temperature for a given nanotube. For 10 nm long (10, 10) CNT the saturated value is $\sim 0.39$. This value is less than 1 (perfect alignment of all the dipoles along the nanotube axis) because perfect alignment along the axis is energetically unfavorable due to the loss of hydrogen bonds.\\
Such alignment of dipoles is very similar to the paraelectric behavior of dipoles under the effect of an aligning field. This led us to propose an analogy with the Langevin theory of paraelectrics. In this description, the interaction between atomic dipoles is negligible and they are randomly oriented at a finite temperature in the absence of aligning electric fields. In the presence of an external electric field $(E_a)$, the dipole moments tend to orient along the field direction. The average polarization per unit volume for a given field at temperature $T$ is given by:  $D=D_0\left( coth \beta -1/\beta\right)$, where  $\beta$ = $\mu_0 E_a/k_B T$ and $D_0=n\mu_0$ ; $n$ being the number of dipoles per unit volume and  $\mu_0$ is the individual dipole moment. To quantify the flow-induced effective field which aligns the confined water dipoles, we propose a similar Langevin function for the net dipole moment $(D)$  of the system of confined water molecules with  $\beta=av/T$ where $a$ is a scaling parameter that describes the proportionality of the effective aligning field with the flow speed $v$:
   \begin{equation}
D=D_0\left[\coth\left(\frac{a v}{T}\right) -\frac{T}{a v}\right]
\end{equation}           
For a given temperature $T$, we have fitted the variation of the average dipole moment per water molecule , projected in the flow direction, with different $v$  to the above Langevin function where $D_0$, the saturation dipole moment and  $a$ are fitting parameters. A good fit shown in Fig.\ref{lang_fit}  implies that a flow-induced field aligning the dipoles is a good description of this effect. This fit also allows us to extract an effective field generated by the flow.
The temporal evolution of the net dipole projection of confined water molecules shown in inset of Fig. 2(b) indicates that the confined molecules prefer aligned states with the dipoles pointing along the CNT axis. The collective flip between parallel (along the flow) and antiparallel (opposite to the flow) aligned states occurs at regular time intervals. (Data shown here is for 10 nm long CNT,  as the length of the CNT increases, this description becomes more accurate, as shown in Fig. S2). The probability distribution of the net dipole projection shown in Fig.\ref{dipole_dist}b depicts that the probability of finding a net dipole moment along +z and -z directions is almost equal without flow. As the flow velocity increases, the probability of finding a net dipole projection aligned with the flow direction increases, as evident from the increasing height of the peak near the net dipole of  $0.4$ and decreasing height of the peak around  $-0.4$ that corresponds to the antiparallel aligned state. Such a variation of probabilities with flow suggests that, on average, confined water molecules spend a longer time in the state with dipoles aligned with the flow direction. This time continues to increase with flow velocities. To quantify this, we compute the average duration for which the confined molecules continuously remain aligned before initiating a successful orientational flip to the opposite alignment. We start by simplifying the dipole trajectory by allowing only three possible values for dipole projections such that a state corresponding to net dipole $\ge 0.2$  is assigned a dipole projection of 1 (parallel to the flow), $\le -0.2$ is assigned as $-1$ (antiparallel to flow), and anything between  $0.2 $ to $-0.2$  is assigned as 0 (transition state). The variation of the average dwell time of parallel and antiparallel alignment with the flow velocity extracted from this modified trajectory is shown in Fig.\ref{dwelltime}a. It clearly demonstrates that the average dwell time for the parallel aligned state increases and the dwell time  for the antiparallel aligned states  decreases with increasing flow velocity.

\begin{figure}
\includegraphics[width=\linewidth]{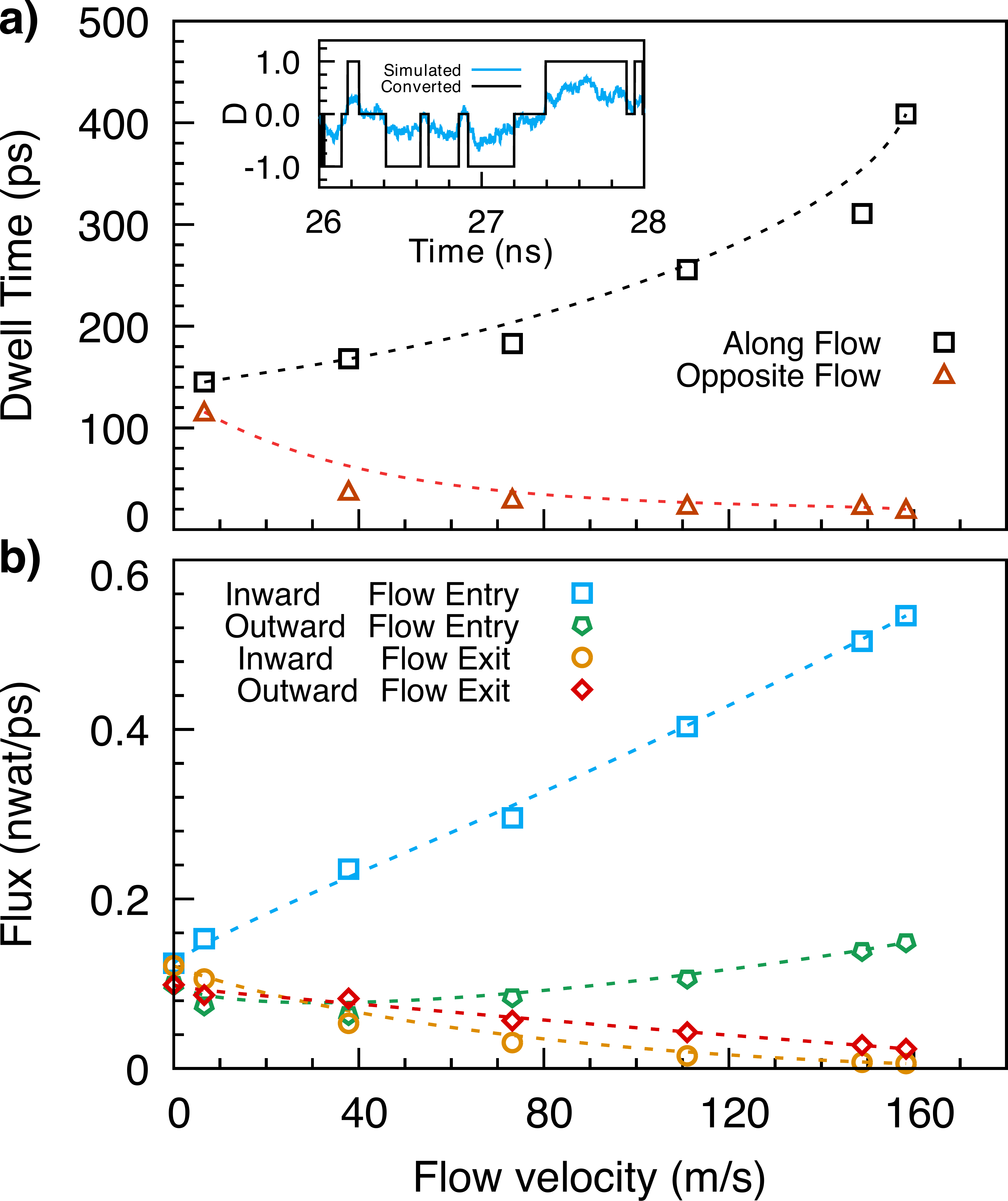}
\caption{   {\bf a)}The variation of the average dwell time for both alignment states with the flow velocity. The dwell time increases with the flow velocity for the parallel alignment and decreases for the state of antiparallel alignment. The inset shows the construction of the simplified dipole trajectory described in the text. {\bf b)} The number of water molecules entering from the bulk into the CNT cavity with their dipole moments pointing inward or outward. More water molecules enter with dipole pointing inward to the nanotube from both ends at equilibrium. As the flow velocity increases from zero, the flux from the flow-entry end increases while the flux from the opposite end decreases.(Lines are the guide to the eye only). }
\label{dwelltime}
\end{figure}
To understand how flow impacts the dwell time, we begin by analyzing the reorientation mechanism in equilibrium conditions~\cite{laage2012reorientation, laage2012water}. Dipole reorientation inside narrow CNTs, where the dipoles are aligned in a single file manner, is initiated through defect nucleation at one of the CNT ends~\cite{doi:10.1021/nn800182v}. When a water molecule enters the CNT such that its dipole is pointing in the opposite direction to those of the confined water dipoles, it is characterized as an orientational defect. When such a defect moves through the confined water chain, it leaves behind flipped individual dipoles. A successful collective flip of the dipole moments occurs when a defect entering from one end exits from the other end. A similar reorientation mechanism can be extended for the wider CNT considered here. First, we characterize the dipole moments of water molecules that enter the CNT. We separated the entry events based on dipole moment of the entering water molecules. A strong preference was observed for water molecules to enter with dipoles pointing inwards (dipole projection $> 0$ for the left end, $< 0 $ for the right end) in equilibrium condition. Of all the water molecules entering inside the CNT cavity from both ends, 55\% of water molecules have dipole pointing inward while only 45\% have dipole pointing outwards. Moreover, the average interaction energy of a water molecule just before entry with its dipole vector pointing inward is  $-9.798$ kcal/mol and  $-9.391$  kcal/mol for outward-pointing dipole vector of entering water molecules. This preference arises due to distinct hydrogen bonding environments near the entrance of the nanotube for the two dipole orientations, as highlighted in Fig. ~\ref{simbox}(c, d). Due to different surrounding environments, the average number of hydrogen bonds for an inward-pointing dipole is  $3.8 $ as compared to $3.4$ for the outward-pointing dipoles, thus showing energetics to be responsible for the preferred entry of molecules with inward-pointing dipoles.\\

Favorable bias for the entry of inward-pointing water molecules and defect-assisted orientational flip mechanism are the key ingredients to understand the emergence of net alignment due to flow. In the absence of flow, the influx from both ends of the CNT is equal; hence the probability of defect nucleation to initiate the collective dipole flip is the same for both the net alignments. This explains the similar dwell times for both the alignments. However, during the flow, a larger number of water molecules enter from the flow-entry edge of CNT as compared to the flow-exit edge. As there is a strong preference for water molecules to enter with their dipoles pointing inward, more water molecules will enter through the flow-entry end with dipole pointing in the flow direction [Fig. 4b]. These water molecules do not destroy the alignment of the dipoles inside the CNT if the alignment is in the flow direction. However, a smaller number of water molecules also go inside the CNT with their dipole moments pointing outward and serve as orientational defects if the dipolar alignment inside the CNT is in the direction of the flow. The situation is reversed if the alignment of the dipoles inside the CNT is opposite to the flow direction: in that case, the larger number of water molecules entering with dipole moments pointing inwards act as defects that can lead to a collective flip of the dipolar alignment. Hence, the probability of a reorientational flip is higher for antiparallel alignment. The net effect of preferred orientation at the flow-entry edge is to increase the probability of defect nucleation for antiparallel alignment. At the opposite end, water molecules prefer to enter with their dipoles pointing opposite to the flow (inward). Such entry events are likely to initiate flips for a parallel aligned state. However, being the flow-exit edge, the influx from this end decreases with the flow velocity resulting in a reduced probability of flipping for the parallel aligned state [Fig. 4b]. 

To summarize, we have demonstrated that a net dipole alignment in the flow direction emerges for the water molecules flowing through a (10, 10) CNT and the dipole alignment increases with the flow velocity. The net alignment arises due to the asymmetry in the dipole orientation with which water molecules enter the nanotube. The preferential entry of water molecules with dipole pointing inwards decreases (increases) the probability of orientational flips when the confined water dipoles are aligned parallel (antiparallel) to the flow direction. This makes the time-averaged net dipole projection in the steady state to point in the flow direction. This observation presents a unique example where confinement produces entirely different behavior from the bulk counterparts. 

The correlation between dipole alignment and flow rate can be exploited for various nanofluidic devices as it gives us a way to control the net dipole moment via external pressure. This observation also raises an important question: is dipolar alignment necessary for the ultrafast transport which has been observed in narrow nanopores? The answer to this question requires more detailed investigation. The net dipole moment of water inside a carbon nanotube can polarize it and this effect can be used to engineer a potential difference across nanotube ends~\cite{zhao_jacs_2009}. As demonstrated here, such polarization can be obtained inside wider nanotubes that confine more significant dipole moments per unit length. It will potentially lead to more significant potential differences across the nanotube ends. As the flow modulates dipole relaxation, this also offers a possibility of making devices where the flow velocity of the water controls the dielectric constant.


\begin{acknowledgments}
HK would like to thank National PARAM Supercomputing Facility at C-DAC, India for computational support. PKM, CD and AKS thank DST, India for financial support.  
\end{acknowledgments}

%
\end{document}